\documentclass[preprint2]{aastex62}
\usepackage{amsmath,amstext}
\usepackage[version=4]{mhchem}
\usepackage{ragged2e}
\usepackage{apjfonts} 
\usepackage{gensymb}
\usepackage{multirow}
\usepackage{tikz}
\usepackage{graphics}
\usepackage[figure,figure*]{hypcap}
\usepackage[para,online,flushleft]{threeparttable}
\usepackage{comment}
\usepackage{tablefootnote}
\usepackage{subscript}

\makeatletter
\newcounter{reaction}
\renewcommand\thereaction{R\arabic{reaction}}
\@addtoreset{reaction}{chapter}
\newcommand\reactiontag%
{\refstepcounter{reaction}\tag{\thereaction}}
\newcommand\reaction@[2][]%
{\begin{equation}\ce{#2}%
\ifx\@empty#1\@empty\else\label{#1}\fi%
\reactiontag\end{equation}}
\newcommand\reaction@nonumber[1]%
{\begin{equation*}\ce{#1}\end{equation*}}
\newcommand\reaction%
{\@ifstar{\reaction@nonumber}{\reaction@}}
\makeatother

\graphicspath{{./}{figures/}}

\received{XXX}
\revised{XXX}
\accepted{XXX}
\submitjournal{ApJ}

\shorttitle{Lifetime of Nitrogen}
\shortauthors{Hu \& Delgado}

\begin{document}

\title{Stability of Nitrogen in Planetary Atmospheres in Contact with Liquid Water\footnote{$\copyright$2019. California Institute of Technology. Government sponsorship acknowledged.}}

\correspondingauthor{Renyu Hu}
\email{renyu.hu@jpl.nasa.gov}

\author[0000-0003-2215-8485]{Renyu Hu}
\affiliation{Jet Propulsion Laboratory \\
California Institute of Technology \\
Pasadena, CA 91109, USA}
\affiliation{Division of Geological and Planetary Sciences \\
California Institute of Technology \\
Pasadena, CA 91125, USA}

\author{Hector Delgado Diaz}
\affiliation{Jet Propulsion Laboratory \\
California Institute of Technology \\
Pasadena, CA 91109, USA}

\begin{abstract}

Molecular nitrogen is the most commonly assumed background gas that supports habitability on rocky planets. Despite its chemical inertness, nitrogen molecule is broken by lightning, hot volcanic vents, and bolide impacts, and can be converted into soluble nitrogen compounds and then sequestered in the ocean. The very stability of nitrogen, and that of nitrogen-based habitability, is thus called into question. Here we determine the lifetime of molecular nitrogen vis-\`a-vis aqueous sequestration, by developing a novel model that couples atmospheric photochemistry and oceanic chemistry. We find that \ce{HNO}, the dominant nitrogen compounds produced in anoxic atmospheres, is converted to \ce{N2O} in the ocean, rather than oxidized to nitrites or nitrates as previously assumed. This \ce{N2O} is then released back into the atmosphere and quickly converted to \ce{N2}. We also find that the deposition rate of \ce{NO} is severely limited by the kinetics of the aqueous-phase reaction that converts \ce{NO} to nitrites in the ocean. Putting these insights together, we conclude that the atmosphere must produce nitrogen species at least as oxidized as \ce{NO2} and \ce{HNO2} to enable aqueous sequestration. The lifetime of molecular nitrogen in anoxic atmospheres is determined to be $>1$ billion years on temperate planets of both Sun-like and M dwarf stars. This result upholds the validity of molecular nitrogen as a universal background gas on rocky planets.

\end{abstract}


\section{Introduction} \label{sec:intro}

Nitrogen is the bulk constituent of Earth's atmosphere and a common constituent of the atmospheres of rocky planets in the Solar System. The universality of nitrogen has been extended to extrasolar rocky worlds, as molecular nitrogen (\ce{N2}) is generally assumed as the background gas in the atmosphere. The standard picture of habitable planets of stars \citep{Kasting:1993zz} posits that climate and geologic processes on rocky planets regulate the abundance of atmospheric \ce{CO2} to maintain a surface temperature that is consistent with liquid water oceans -- but an often overlooked ingredient of this picture is a constant, approximately 1 bar, \ce{N2}-dominated background atmosphere.

The climate-maintaining effect of \ce{N2} primarily stems from its higher volatility than \ce{CO2} or \ce{H2O}. As the partial pressure of \ce{CO2} is controlled by the silicate weathering cycle \citep{Walker1981}, and that of \ce{H2O} is controlled by the surface temperature, the partial pressure of \ce{N2} is not a direct function of any climatological parameters. Having a sizable \ce{N2} atmosphere, therefore, alleviates the sensitivity of the planetary climate to subtle changes in forcings, and thus widens the semi-major axis ranges in which the planet can be habitable \citep{Vladilo2013}. No habitable climate can be found if the partial pressure of \ce{N2} is less than 0.015 bar \citep{Vladilo2013}. The actual lower-limit may be even higher, as it is later found that \ce{N2} as a non-condensable gas maintains the cold trap of the middle atmosphere and prevents water from loss to space \citep{Wordsworth2013}. While \ce{N2} is not a greenhouse gas, it causes a warming effect on climate via pressure broadening of \ce{CO2} and \ce{H2O} absorption features \citep{Goldblatt2009}.

Due to its strong triple bond, \ce{N2} is very close to being chemically inert in the atmosphere. The processes that can break \ce{N2} are peculiar \citep{Mancinelli1988}: on today's Earth it is primarily performed by microbes, and before the rise of nitrogen-fixation microbes it is done in energetic events including lightning \citep{Yung1979,Kasting1981,Navarro1998,Wong2017}, bolide impact \citep{Mckay1988}, and also hot volcanic vents \citep{Mather2004}. The immediate product of the ``atmospheric nitrogen fixation'' is \ce{NO}, and the \ce{NO} is then converted to \ce{HNO3} in oxygen-rich atmospheres and to \ce{HNO} in oxygen-poor ones \citep{Kasting1981,Wong2017}. It has been suggested that the \ce{HNO} is then converted to \ce{NO2-} and \ce{NO3-} in the ocean \citep{Mancinelli1988,Summers2007}. As such, \ce{NO} produced in the atmosphere eventually becomes nitrites and nitrates. The entire 1-bar \ce{N2}-dominated atmosphere could be sequestered in the ocean as nitrites and nitrates -- thus creating a potential problem for the stability of a nitrogen-dominated atmosphere in contact with liquid water oceans.

We are therefore motivated to determine the lifetime of \ce{N2} -- and thus that of \ce{N2}-based habitability -- on a habitable exoplanet. We focus on anoxic planets without life, because microbes would be able to harvest the nitrites and nitrates in the oceans, reduce them to \ce{N2} or \ce{N2O}, and restore the \ce{N2} stability. Without life, the formation of nitrites and nitrates may well be mostly one-way and become long-term losses of nitrogen. In this paper we calculate the kinetic timescale of this process. We first study the fate of \ce{HNO}, the dominant nitrogen compound produced in anoxic atmospheres, when it is deposited into the ocean. We show that \ce{HNO} does not lead to nitrogen sequestration but rather formation of \ce{N2O} (Section \ref{sec:fate}). We then present a novel model that couples an atmosphere photochemistry model \citep{Hu2012,Hu2013} and an ocean aqueous-chemistry model, so that the rates of transfer between the atmosphere and the ocean can be self-consistently calculated (Section \ref{sec:model}). Using the coupled model we determine the lifetime of \ce{N2} in anoxic atmospheres on temperate planets of Sun-like and M dwarf stars (Section \ref{sec:results}). We discuss the implications of our findings in Section \ref{sec:discussion} and conclude in Section \ref{sec:conclusion}. 

\section{Aqueous Chemistry of \ce{HNO} on Planets} \label{sec:fate}

\subsection{Aqueous-Phase Reactions and Kinetic Rates}

\ce{HNO} is the main atmospheric product of nitrogen compounds under anoxic conditions, and its fate in the ocean has not been clarified. The aqueous chemistry of \ce{HNO}, and its conjugate base \ce{NO-}, is peculiar because the ground state of \ce{HNO} is a singlet while that of \ce{NO-} is a triplet. This makes the deprotonation reaction to proceed as the forward direction of
\reaction{HNO + OH- <=> NO- + H2O \label{R:HNOD}}
a slow, second-order reaction \citep{miranda2005chemistry}. Under the pH conditions relevant to planets, most of the dissolved \ce{HNO} exists in the form of \ce{HNO}. We note that the excited state \ce{HNO} is a triplet and it quickly dissociates to \ce{NO-}. The transition to the excited state, however, is spin forbidden and has not been observed in experiments.

Dissolved \ce{HNO} can be removed by rapid dehydrative dimerization
\reaction{HNO + HNO -> N2O + H2O \label{R:HNOHNO}} 
with its rate constant determined by the flash photolysis technique \citep{shafirovich2002nitroxyl}. 

\ce{NO-} is rapidly oxidized to nitrate when free oxygen is available
\reaction{NO- + O2 -> ONOO- -> NO3- \label{R:NOO2}}
or polymerized by NO via
\reaction{NO- + NO <=> N2O2- \label{R:NONO1}}
\reaction{N2O2- + NO -> N3O3 \label{R:NONO2}}
\reaction{N3O3- -> N2O + NO2- \label{R:NONO3}}
The polymerization can also start from \ce{HNO}
\reaction{HNO + NO <=> HN2O2 \label{R:HNONO1}}
\reaction{HN2O2 + NO -> HN3O3 \label{R:HNONO2}}
\reaction{HN3O3 -> N2O + HNO2 \label{R:HNONO3}}
Both polymerization reactions eventually form \ce{N2O} and nitrite, and their rate constants have been measured using pulse radiolysis and NO-rich fluids \citep{gratzel1970pulse,seddon1973pulse}. These polymerization pathways have been adopted as the pathway to convert \ce{HNO} to nitrite and nitrate in planetary oceans \citep{Mancinelli1988,Summers2007,Wong2017}.

In summary, the removal pathways of \ce{HNO} in the aqueous phase are dehydrative dimerization (Reaction \ref{R:HNOHNO}), deprotonation (Reaction \ref{R:HNOD}) followed by either oxidation (Reaction \ref{R:NOO2}) or polymerization (Reactions \ref{R:NONO1}-\ref{R:NONO3}), and direct polymerization (Reactions \ref{R:HNONO1}-\ref{R:HNONO3}). Relevant rate constants are tabulated in Table \ref{tab:rate}. 

\begin{table}[h!]
\centering
\begin{tabular}{ll} 
 \hline\hline
 Reaction & Rate Constant \\
 \hline
 \ref{R:HNOD} forward & $5\times10^4$ M$^{-1}$ s$^{-1}$ \\
 \ref{R:HNOD} reverse & $1.2\times10^2$ s$^{-1}$  \\
 \ref{R:HNOHNO} & $8\times10^6$ M$^{-1}$ s$^{-1}$ \\
 \ref{R:NOO2} & $4\times10^9$ M$^{-1}$ s$^{-1}$ \\
 \ref{R:NONO1} forward & $2\times10^9$ M$^{-1}$ s$^{-1}$ \\
 \ref{R:NONO1} reverse & $3\times10^4$ s$^{-1}$ \\
 \ref{R:NONO2} & $3\times10^6 $ M$^{-1}$ s$^{-1}$ \\
 \ref{R:HNONO1} forward & $2\times10^9$ M$^{-1}$ s$^{-1}$ \\
 \ref{R:HNONO1} reverse & $8\times10^6$ s$^{-1}$ \\
 \ref{R:HNONO2} & $8\times10^6$ M$^{-1}$ s$^{-1}$ \\
 \ref{R:NO} & $2\times10^8$ M$^{-1}$ s$^{-1}$ \\
 \ref{R:NO2} & $1\times10^8$ M$^{-1}$ s$^{-1}$ \\
 \hline\hline
\end{tabular}
\caption{Rate constants for \ce{HNO}, \ce{NO}, and \ce{NO2} reactions in the aqueous phase. Compiled from \cite{miranda2005chemistry} and \cite{lee1984atmospheric}. The rate constants are provided at the room temperature.}
\label{tab:rate}
\end{table}

\subsection{Reaction Rates under Planetary Conditions}

Using the kinetic constants from experiments, we calculate the reaction rates of the \ce{HNO} removal pathways under typical planetary conditions.

After deprotonation, \ce{NO-} can be either oxidized (Reaction \ref{R:NOO2}) or polymerized (Reactions \ref{R:NONO1}-\ref{R:NONO3}). We first compare the two sub-pathways. The rate of Reaction (\ref{R:NOO2}) is
\begin{equation}
    R_{\ref{R:NOO2}} = k_{\ref{R:NOO2}}\ce{[NO^{-}][O_2]},
\end{equation}
and the overall rate of Reactions (\ref{R:NONO1}-\ref{R:NONO3}) is
\begin{equation}
    R_{\ref{R:NONO1}-\ref{R:NONO3}} = k_{\ref{R:NONO1}f}\ce{[NO^{-}][NO]}\frac{k_{\ref{R:NONO2}}\ce{[NO]}}{k_{\ref{R:NONO1}r}+k_{\ref{R:NONO2}}\ce{[NO]}},
\end{equation}
where the additional $f$ and $r$ in the subscript denote the rate constant of the forward and the reverse directions, respectively, and quantities in \ce{[X]} denote the concentration of the species \ce{X} in the aqueous phase, usually in the unit of M (i.e., mole per liter). The reaction rate $R$ has the unit of M s$^{-1}$.

\begin{table}[h!]
\centering
\begin{tabular}{lll} 
 \hline\hline
 Species & Typical & Upper Limit \\
 \hline
 $f_{\ce{NO}}$ & $4\times10^{-11}$ & $3\times10^{-6}$ \\
 \ce{[NO]} & $8\times10^{-14}$ M & $6\times10^{-9}$ M \\
  $f_{\ce{O2}}$ & $2\times10^{-15}$ & $1\times10^{-8}$ \\
 \ce{[O2]} & $2\times10^{-18}$ M & $1\times10^{-11}$ M \\
  $f_{\ce{HNO}}$ & $2\times10^{-11}$ & $2\times10^{-10}$ \\
 \ce{[HNO]} & $2\times10^{-10}$ M & $2\times10^{-9}$ M \\
 \hline\hline
\end{tabular}
\caption{Typical and anoxic upper-limit concentrations for evaluating and comparing the reaction rates of \ce{HNO} removal pathways. For each gas (\ce{X}), the mixing ratio at the bottom of the atmosphere ($f_{\ce{X}}$) and the concentration in the surface ocean (\ce{[X]}) are provided. These quantities are consistent with the converged photochemistry models shown in Section \ref{sec:results}.}
\label{tab:con}
\end{table}

For a typical anoxic condition, $[\ce{NO}]\sim8\times10^{-14}$ M and $[\ce{O2}]\sim2\times10^{-18}$ M (Table \ref{tab:con}). These quantities are from the atmospheric photochemistry models under terrestrial lightning rates (Section \ref{sec:results}) and have factored in the Henry's law constants for respective gases. When the lightning rate is very high (i.e., $100\times$ the terrestrial rate), the upper limits are $[\ce{NO}]\sim6\times10^{-9}$ M and $[\ce{O2}]\sim10^{-11}$ M. Note that these upper limits do not include the oxygen-rich scenarios that would be produced on planets of M dwarf stars (see Section \ref{sec:results}).

Based on these concentrations, $R_{\ref{R:NONO1}-\ref{R:NONO3}} = 10^{-15}\sim10^{-10} [\ce{NO-}]$ M s$^{-1}$ and $R_{\ref{R:NOO2}} = 10^{-8}\sim4\times10^{-2} [\ce{NO-}]$ M s$^{-1}$. Therefore, even under the anoxic conditions, $R_{\ref{R:NOO2}}\gg R_{\ref{R:NONO1}-\ref{R:NONO3}}$, and the same is true for oxygen-rich conditions. The overall rate of the removal path starting with deprotonation (Reaction \ref{R:HNOD}) is thus
\begin{equation}
    R_{\ref{R:HNOD}} = k_{\ref{R:HNOD}f}\ce{[HNO][OH^{-}]}\frac{k_{\ref{R:NOO2}}[\ce{O2}]}{k_{\ref{R:HNOD}r}+k_{\ref{R:NOO2}}[\ce{O2}]}.
\end{equation}

The rate of dehydrative dimerization (Reaction \ref{R:HNOHNO}) is
\begin{equation}
    R_{\ref{R:HNOHNO}} = k_{\ref{R:HNOHNO}}\ce{[HNO][HNO]},
\end{equation}
and the overall rate of direct polymerization (Reactions \ref{R:HNONO1}-\ref{R:HNONO3}) is
\begin{equation}
    R_{\ref{R:HNONO1}-\ref{R:HNONO3}} = k_{\ref{R:HNONO1}f}\ce{[HNO][NO]}\frac{k_{\ref{R:HNONO2}}\ce{[NO]}}{k_{\ref{R:HNONO1}r}+k_{\ref{R:HNONO2}}\ce{[NO]}}.
\end{equation}

Under typical and limiting anoxic conditions, $[\ce{HNO}]\sim2\times10^{-10}-2\times10^{-9}$ M (Section \ref{sec:results}). For a neutral pH, we estimate $R_{\ref{R:HNOD}}=7\times10^{-23}\sim3\times10^{-15}$ M s$^{-1}$, $R_{\ref{R:HNOHNO}}=3\times10^{-13}\sim3\times10^{-11}$ M s$^{-1}$, and $R_{\ref{R:HNONO1}-\ref{R:HNONO3}}=3\times10^{-27}\sim10^{-16}$ M s$^{-1}$. Comparing the three rates, we have $R_{\ref{R:HNOHNO}} \gg R_{\ref{R:HNOD}} > R_{\ref{R:HNONO1}-\ref{R:HNONO3}}$. $R_{\ref{R:HNOD}}$ is proportional to the concentration of \ce{OH-} in the ocean, and for $R_{\ref{R:HNOD}}$ to be greater than $R_{\ref{R:HNOHNO}}$, the ocean must be highly alkaline with pH > 11. Such a pH value is well higher than the pH of Earth's ocean currently or in the Archean \citep{halevy2017geologic,krissansen2018constraining}. Therefore, under anoxic conditions relevant for planetary atmospheres, dehydrative dimerization (Reaction \ref{R:HNOHNO}) is the dominant removal pathway of \ce{HNO} deposited in the ocean.

Under oxygen-rich conditions, including in the oxygen-rich atmospheres produced by \ce{CO2} photolysis on planets of M dwarf stars (see Section \ref{sec:results}), little \ce{HNO} is produced in the atmosphere, and thus the dissolved concentration is very small. In this case, the rate of Reaction (\ref{R:HNOHNO}) is very small, and deprotonation followed by oxidation (Reactions \ref{R:HNOD} and \ref{R:NOO2}) dominates. However, that \ce{HNO} oxidation pathway is still not important to the overall removal flux of nitrogen, because little \ce{HNO} is produced in the atmosphere in the first place.

\subsection{Consistency with \cite{Summers2007}}

The main finding of this section is that under planetary conditions the deposited \ce{HNO} in the ocean does not mainly become nitrite or nitrate. This finding might be perceived as contradictory to the experimental result of \cite{Summers2007}, where a gas mixture of \ce{CO2} and \ce{N2} with 1\% \ce{NO} and 1\% \ce{CO} in contact with liquid water was irradiated by ultraviolet light. \cite{Summers2007} found that nitrate and nitrite to a lesser extent were formed and the \ce{NO} was depleted in approximately 1 hour. A smaller amount of \ce{N2O} was also produced. The interpretation was that \ce{HNO} was formed and dissolved, and Reactions (\ref{R:NONO1}-\ref{R:NONO3}) or Reactions (\ref{R:HNONO1}-\ref{R:HNONO3}) took place dominantly in the system.

The experimental result of \cite{Summers2007} is consistent with our model of the kinetics of \ce{HNO} aqueous chemistry, as it showcases the outcome from a \ce{NO}-rich fluid. The experimental vessel was filled to a pressure of approximately 1 bar, which means that in equilibrium $[\ce{NO}]\sim2\times10^{-5}$ M. Therefore, the fluid was more \ce{NO}-rich than planetary oceans by orders of magnitude. Applying this concentration and re-evaluating all reaction rates in this section, we find that the rate of Reaction (\ref{R:HNOD}) followed by Reactions (\ref{R:NONO1}-\ref{R:NONO3}) is $2\times10^{-3}[\ce{HNO}]$ M s$^{-1}$, the rate of Reactions (\ref{R:HNONO1}-\ref{R:HNONO3}) is $8\times10^{-1}[\ce{HNO}]$ M s$^{-1}$, and the rate of Reaction (\ref{R:HNOHNO}) is $8\times10^{6}[\ce{HNO}]^2$ M s$^{-1}$. The concentration of \ce{HNO} in the system is unknown, but \ce{NO} has a lifetime of 1 hour and yet [\ce{HNO}] has a lifetime of at most $\sim1$ s. As an upper limit, we assume that \ce{HNO} is the only intermediary in the removal of \ce{NO} and that all \ce{HNO} in the system (a 110-ml gas cell) is in the aqueous phase (15-ml water, \cite{Summers2007}). We estimate $[\ce{HNO}]<7\times10^{-7}$ M. Together, we find that even at this upper limit, the reaction rate of direct polymerization (Reactions \ref{R:HNONO1}-\ref{R:HNONO3}) is on the same order of magnitude as the reaction rate of dehydrative dimerization (Reaction \ref{R:HNOHNO}). In reality, the concentration of \ce{HNO} should be smaller and polymerization becomes the dominant pathway, with the \ce{N2O}-producing dimerization the secondary pathway. This is what was seen in the experiment, and our kinetic model is thus consistent with the experiment.

\subsection{The Fate of \ce{HNO} in Planetary Oceans}

To summarize, the analysis in this section shows that under planetary conditions most of the deposited \ce{HNO} undergoes dehydrative dimerization, and becomes \ce{N2O}. The dehydrative dimerization is kinetically favored over oxidization to nitrate or polymerization to nitrite by at least four orders of magnitude under anoxic conditions, and in most cases, by ten orders of magnitude. 

The insight we obtain here by evaluating the kinetic rates of \ce{HNO} removal pathways clarifies the fate of \ce{HNO} produced in anoxic atmospheres and deposited in the oceans. Models of the atmospheric evolution for Earth and planets have assumed that the \ce{HNO} would quickly become nitrite and nitrate in the ocean \citep{Mancinelli1988,Wong2017,laneuville2018earth,ranjan2019nitrogen}. The experimental basis for this early assumption was the pulse radiolysis experiments for Reactions \ref{R:NONO1}-\ref{R:NONO3} and \ref{R:HNONO1}-\ref{R:HNONO3} \citep{gratzel1970pulse,seddon1973pulse} and the experiment of \cite{Summers2007}. These experiments used \ce{NO}-rich fluids, and thus to apply their results one must evaluate the implied kinetic rates for reasonable planetary conditions and compare with other potential reaction pathways. Here we show that for anoxic atmospheres, dehydrative dimerization is the dominant pathway, and for oxygen-rich atmospheres, deprotonation followed by oxidation is the dominant pathway. {These results are also testable by experiments in the laboratory.}

It is therefore reasonable to consider Reaction (\ref{R:HNOHNO}) the sole reaction of \ce{HNO} in the aqueous phase. The produced \ce{N2O}, because of its low solubility, is released to the atmosphere and eventually photolyzed to become \ce{N2}. The formation of \ce{HNO} in the atmosphere is thus not an effective path toward nitrite or nitrate, and does not lead to sequestration of molecular nitrogen in the aqueous phase.

\section{Coupled Atmosphere-Ocean Model} \label{sec:model}

We develop an ocean chemistry module and couple it with the atmospheric photochemistry model of \cite{Hu2012,Hu2013} to determine the lifetime of \ce{N2} in anoxic atmospheres in contact with liquid-water oceans. The photochemistry model has been validated by computing the atmospheric compositions of present-day Earth and Mars, as the outputs agreed with the observations of major trace gases in Earth's and Mars' atmospheres \citep{hu2013thesis}. The model includes a comprehensive reaction network for O, H, C, N, and S species including sulfur and sulfuric acid aerosols, and its applications to simulating anoxic atmospheres and maintaining the redox flux balance of the atmosphere and the ocean have been well-documented \citep{James2018} and compare well with other photochemical models \citep{Gao2015,Harman2018}.

For this work, we choose to simulate a 1-bar atmosphere of 95\% \ce{N2} and 5\% \ce{CO2}, as this kind of anoxic atmosphere is akin to the \ce{O2}-poor and \ce{CO2}-rich environment of the Archean Earth, and is often adopted as the archetype for anoxic exoplanet atmospheres \citep[e.g.,][]{TIAN2014,domagal2014abiotic,Harman2015}. We assume a surface temperature of 288 K and a stratospheric temperature of 200 K and include volcanic outgassing of \ce{CO}, \ce{H2}, \ce{SO2}, and \ce{H2S} in the same way as in \cite{James2018}. We use the entire reaction network of the atmospheric photochemistry model of \cite{Hu2012,Hu2013}, except the organic compounds that have more than two carbon atoms and their reactions. The outgassing rate adopted here is not high enough to produce a \ce{H2SO4} aerosol layer in the atmosphere.

We include both a Sun-like star and an M dwarf star as the parent star. For the M dwarf star, we use GJ~876 as the representing case and apply its measured spectrum in the ultraviolet \citep{France2016} in the photochemistry model.

To simulate the effect of atmospheric nitrogen fixation, we start from the terrestrial production rate of \ce{NO} by lightning, $6\times10^8$ cm$^{-2}$ s$^{-1}$ \citep{schumann2007global}. Changing the main oxygen donor from \ce{O2} to \ce{CO2} and \ce{H2O} would lead to approximately one-order-of-magnitude less \ce{NO}, but the lightning rate also depends on how convective the atmosphere is \citep{Wong2017,Harman2018}. Besides, bolide impacts and hot volcanic vents may also contribute substantially to the source of \ce{NO} \citep{Mckay1988,Mather2004}. We therefore explore the effect of changing \ce{NO} flux by three orders of magnitude from the terrestrial lightning value to cover these varied scenarios. Also, assuming the oxygen comes from \ce{CO2}, each molecule of \ce{NO} produced is accompanied by another molecule of \ce{CO}. We include this conjugate \ce{CO} source in the model, and in this way, no net redox change is introduced to the atmosphere.

\subsection{Ocean Chemistry and Deposition Velocities of Nitrogen Species} \label{sec:depo}

Chemical reactions in the ocean affect the atmospheric photochemistry model by adjusting the rate of gas exchange between the atmosphere and the ocean. Conceptually, the transfer flux from the atmosphere to the ocean can be expressed as $\phi=v_{\rm max}(n-MC/H)=v_{\rm dep}n$ where $v_{\rm max}$ is the maximum deposition velocity and $v_{\rm dep}$ is the effective deposition velocity, $n$ is the number density at the bottom of the atmosphere, $M$ is the concentration at the surface ocean, $H$ is Henry's law constant, and $C$ is a unit conversion factor depending on the definition of Henry's law constant \citep{kharecha2005coupled}. The effective deposition velocity depends on how fast the ocean can ``process'' the deposited gas: if the ocean removes the gas quickly, then $M\rightarrow0$, and $v_{\rm dep}\rightarrow v_{\rm max}$; whereas if the ocean cannot remove the gas, Henry's law equilibrium could be established, and in this case, $M\rightarrow nH/C$ and $v_{\rm dep}\rightarrow0$. $v_{\rm max}$ can be approximated by the speed for the gas to diffuse through laminar layers at the interface between the atmosphere and the ocean, aka. the two-film model \citep{broker1982tracers} and is sensitive to the solubility of the gas, the wind speed, and the temperature \citep{domagal2014abiotic,Harman2015}. For highly soluble species $v_{\rm max}\sim1$ cm s$^{-1}$ and for weakly soluble ones, $v_{\rm max}\sim10^{-4}-10^{-3}$ cm s$^{-1}$.

\begin{table}[h!]
\centering
\begin{tabular}{lc} 
 \hline\hline
 Species & Deposition velocity \\
 & cm s$^{-1}$ \\
 \hline
 \ce{N2O} & 0 \\
 \ce{NO} & calculated iteratively  \\
 \ce{NO2} & calculated iteratively \\
 \ce{NO3} & 1 \\
 \ce{N2O5} & 1 \\
 \ce{HNO} & calculated iteratively \\
 \ce{HNO2} & 1 \\
 \ce{HNO3} & 1 \\
 \ce{HNO4} & 1 \\
 \hline\hline
\end{tabular}
\caption{Effective deposition velocities for nitrogen species.}
\label{tab:depo}
\end{table}

Table \ref{tab:depo} lists the effective deposition velocities for nitrogen species. We do not include any process that removes \ce{N2O} in the ocean, and thus its deposition velocity is zero. For \ce{HNO2} and \ce{HNO3}, the ocean's capacity to store them is vast, and thus we assume that they are permanently lost to the ocean once deposited, and their deposition velocities approach $v_{\rm max}$. \ce{NO3}, \ce{N2O5}, and \ce{HNO4} quickly react or decomposes to \ce{NO3-}, and thus they are also considered permanently lost once deposited. Over geologic timescales the dissolved \ce{NO2-} and \ce{NO3-} can be reduced to \ce{NH4+}, or to \ce{NO}, \ce{N2O}, and \ce{N2} and released back to the atmosphere, by cycling through hydrothermal vents \citep{Wong2017,laneuville2018earth}, and ultraviolet photolysis and reduction by \ce{Fe^{2+}} \citep[e.g.,][]{stanton2018nitrous,ranjan2019nitrogen}. This potential source of gaseous \ce{NO} and \ce{N2O} is not included in the current model since we explore a wide range of NO flux as the boundary condition, and the \ce{N2O} is readily photodissociated in the atmosphere.

\begin{figure}
    \centering
    \plotone{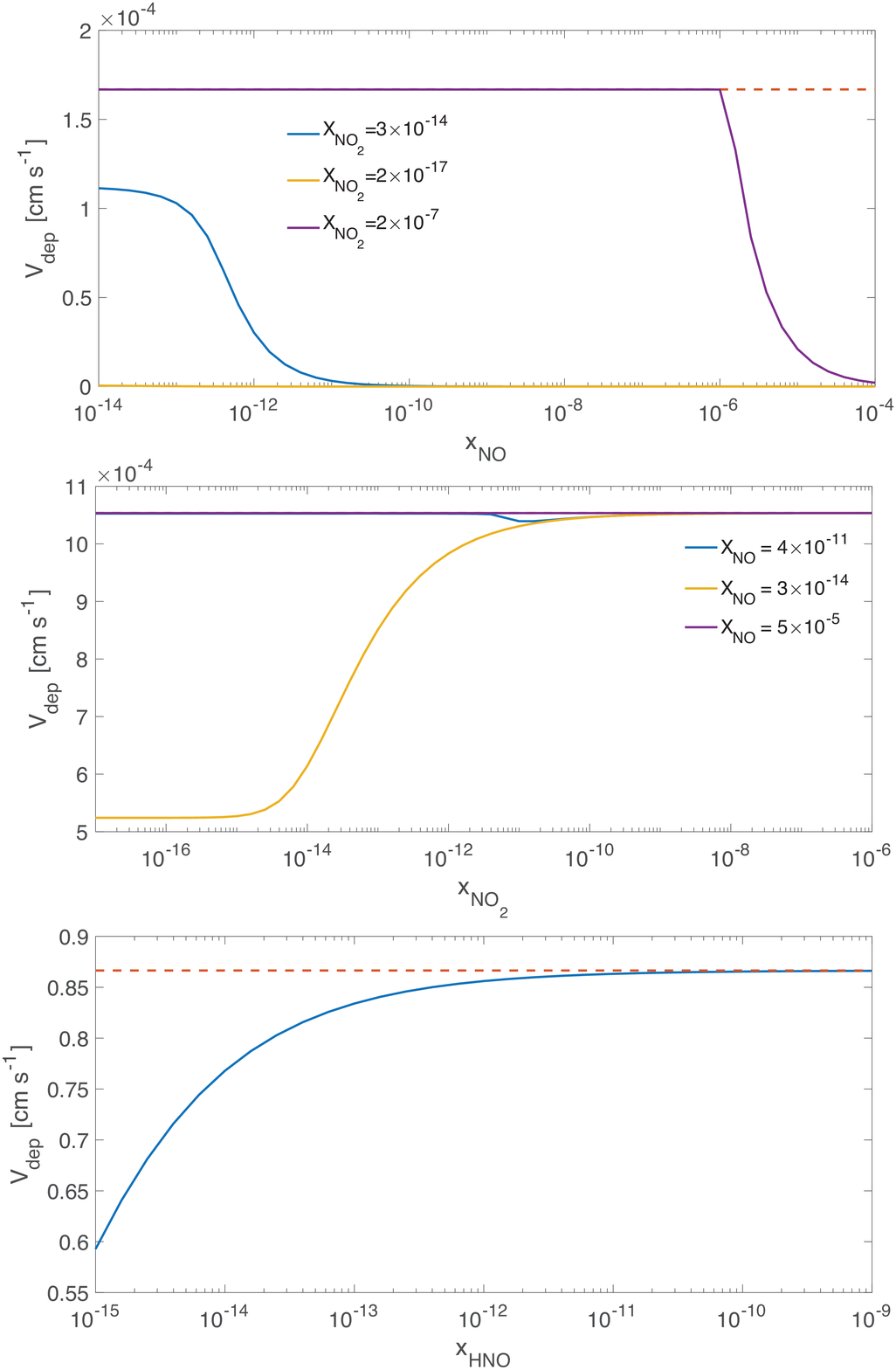}
    \caption{Effective deposition velocities of \ce{NO}, \ce{NO2}, and \ce{HNO} as a function of the partial pressure at the bottom of the atmosphere. The dashed lines show $v_{\rm max}$. The deposition velocities of \ce{NO} and \ce{NO2} depend on the partial pressure of the other gas. For this reason, three cases are shown, with typical (blue lines), low (yellow lines), and high (purple lines) abundances of the other gas. The calculations are performed for a 3-km deep, homogeneous ocean.
    }
    \label{fig:depo}
\end{figure}

For \ce{NO}, \ce{NO2}, and \ce{HNO}, we solve for their concentrations in the ocean, using the rates of Reaction (\ref{R:HNOHNO}) and the following reactions in the aqueous phase:
\reaction{NO + NO2 -> 2NO2- + 2H+  \label{R:NO}}
\reaction{NO2 + NO2 -> NO2- + NO3- + 2H+  \label{R:NO2}}
The rate constants of Reactions (\ref{R:NO}) and (\ref{R:NO2}) are from \cite{lee1984atmospheric} and tabulated in Table \ref{tab:rate}. For each mixing ratio (or partial pressure) of \ce{NO}, \ce{NO2}, and \ce{HNO} at the bottom of the atmosphere, their steady-state concentrations in the ocean can be calculated, assuming homogeneous distribution in the ocean. The results are then expressed in the effective deposition velocities and are shown in Figure \ref{fig:depo}.

Several important observations can be drawn from Figure \ref{fig:depo}. (1) \ce{NO} does not substantially transfer to the ocean unless the mixing ratio of \ce{NO2} is approaching 1 ppm. This is because the removal of \ce{NO} by Reaction (\ref{R:NO}) requires another \ce{NO2}. The conditions for such a large abundance of \ce{NO2} at the surface is rarely achieved. The effective deposition velocity of \ce{NO} can be large when the mixing ratio of \ce{NO} is very small. This however does not imply a substantial transfer flux because the flux is the product of the deposition velocity and the mixing ratio. The deposition flux of \ce{NO} is thus severely limited by the kinetic rate of Reaction (\ref{R:NO}). (2) \ce{NO2} practically deposits at $v_{\rm max}$. Unless the lightning rate is very small, the partial pressure of \ce{NO} is always high enough to effectively remove \ce{NO2} via Reaction (\ref{R:NO}). Even when the mixing ratio of \ce{NO} is indeed very small (see Figure \ref{fig:depo}, middle panel, yellow line), Reaction (\ref{R:NO2}) can efficiently remove the dissolved \ce{NO2} and make the deposition velocity to approach $v_{\rm max}$ for a mixing ratio of \ce{NO2} greater than $10^{-12}$. Since the deposition flux would always be small at the low end of the lightning rate, Figure \ref{fig:depo} indicates that in practice the deposition of \ce{NO2} is always efficient. (3) The deposition of \ce{HNO} is generally quite efficient, with $v_{\rm dep}$ close to $v_{\rm max}$. But as shown in Section \ref{sec:fate}, this deposition leads to a return flux of \ce{N2O} to the atmosphere.

Because the effective deposition velocities depend on the partial pressure at the bottom of the atmosphere, we need to solve the coupled atmosphere-ocean chemistry model iteratively. For each scenario, we typically start with $v_{\rm max}$. Once a steady-state solution is found for the atmospheric chemistry, we use the mixing ratio of \ce{NO}, \ce{NO2}, and \ce{HNO} at the bottom of the atmosphere to calculate their effective deposition velocities. We also add the corresponding return flux of \ce{N2O} as part of the revised boundary conditions. We then relaunch the atmospheric chemistry calculation and find a new steady-state solution. This procedure is repeated until the steady-state mixing ratios of \ce{NO}, \ce{NO2}, and \ce{HNO} no longer change. Typically only a handful of iterations are required. As such, we can found self-consistent solutions that satisfy both the atmosphere and ocean chemistry.

To summarize, the analysis presented so far indicates that the deposition of \ce{NO} or \ce{HNO} cannot be a net sink for molecular nitrogen in the atmosphere, because \ce{NO} does not deposit efficiently and \ce{HNO} deposition leads to a return flux of \ce{N2O}. Therefore, to sequester nitrogen in the ocean, the atmosphere must oxidize nitrogen compounds to at least as oxidized as \ce{NO2} and \ce{HNO2}. With this insight, we will show in Section \ref{sec:results} that this required oxidization is quite slow in anoxic atmospheres and molecular nitrogen is therefore kinetically stable.

\section{Results} \label{sec:results}

\begin{figure}
    \centering
    \plotone{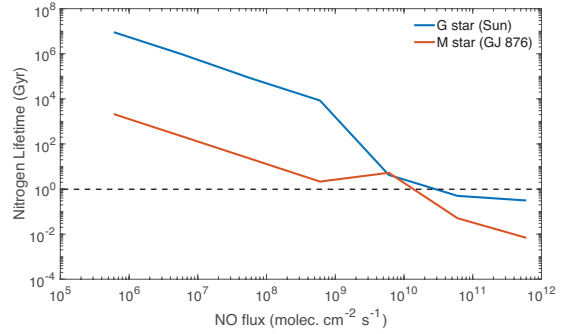}
    \caption{
    Lifetime of molecular nitrogen in planetary atmospheres in contact with liquid water oceans. The dashed line shows 1 billion years for comparison. Lightning in Earth's atmosphere produces a \ce{NO} flux of $6\times10^8$ molecule cm$^{-2}$ s$^{-1}$. The lifetime is well greater than 1 billion years unless the \ce{NO} source flux is particularly strong.}
    \label{fig:lifetime}
\end{figure}


\begin{figure}
    \centering
    \plotone{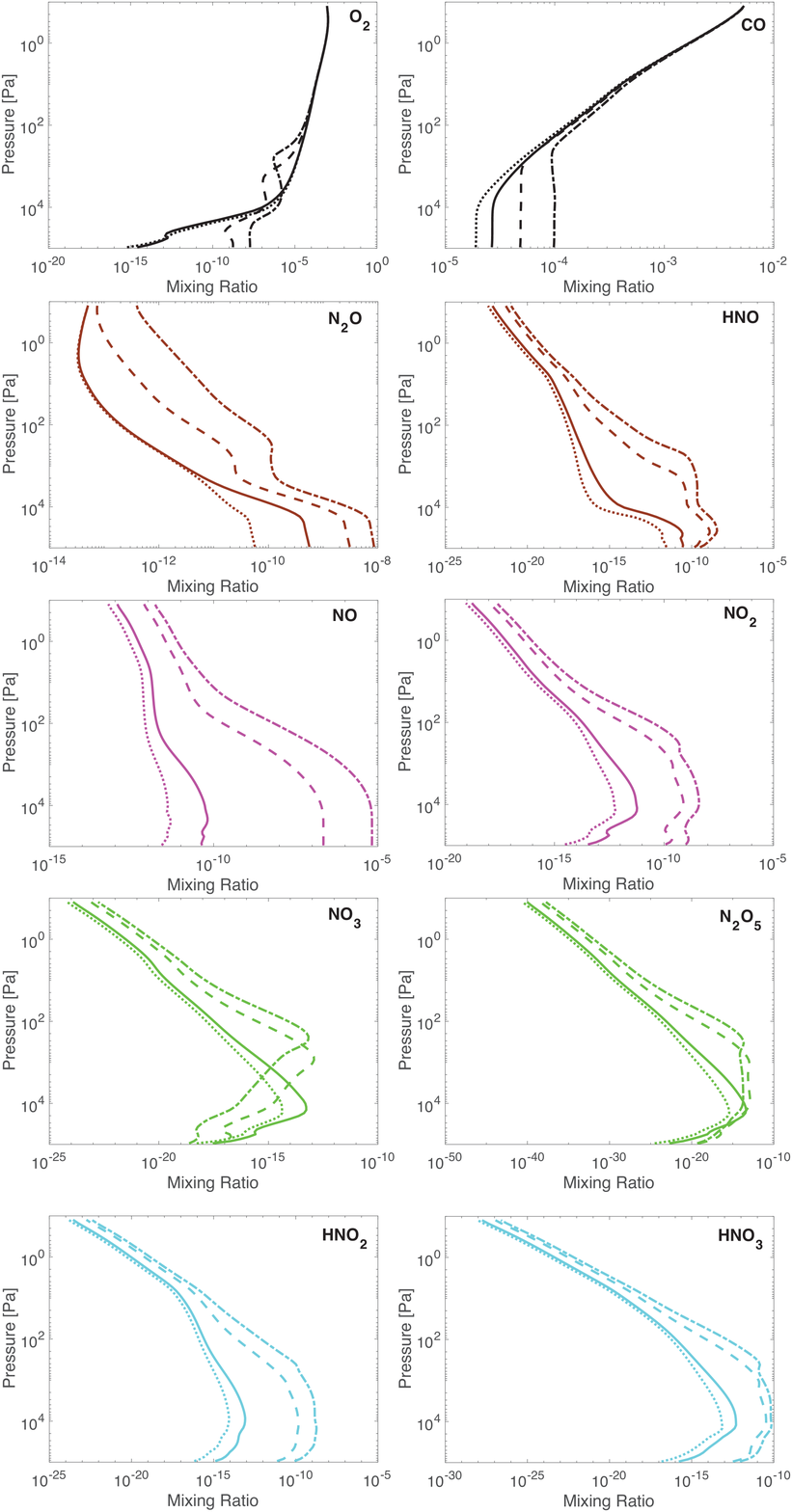}
    \caption{
    The atmospheric abundances of \ce{O2}, \ce{CO}, and main nitrogen species in an \ce{N2}-dominated atmosphere on a temperate rocky planet of a Sun-like star. Note that the horizontal axis of each panel is different. Dotted, solid, dashed, and dash-dot lines are from converged atmosphere-ocean chemistry models with an \ce{NO} flux of $6\times10^7$, $6\times10^8$ (terrestrial value), $6\times10^9$, $6\times10^{10}$ molecule cm$^{-2}$ s$^{-1}$, respectively. The source strength of \ce{NO} has a variety of impact and feedback on the nitrogen chemistry in the atmosphere (see text).}
    \label{fig:sun_detail}
\end{figure}

\begin{figure}
    \centering
    \plotone{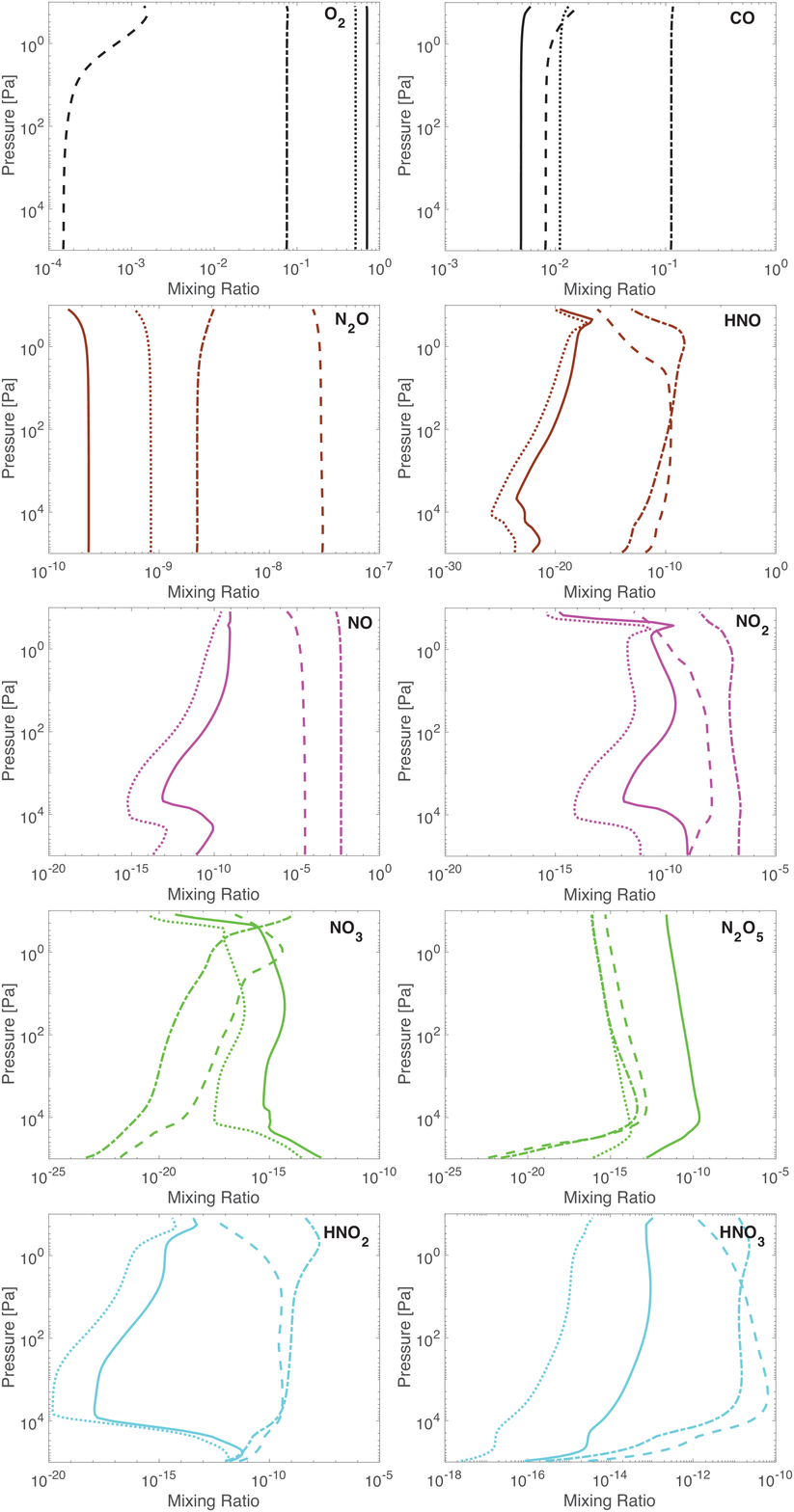}
    \caption{
    The same as Figure \ref{fig:sun_detail} but with GJ~876 as the parent star. The atmosphere becomes \ce{O2}-rich at the steady state due to \ce{CO2} photolysis, and this effect has a strong impact on the nitrogen chemistry (see text).}
    \label{fig:gj876_detail}
\end{figure}

\begin{figure}
    \centering
    \plotone{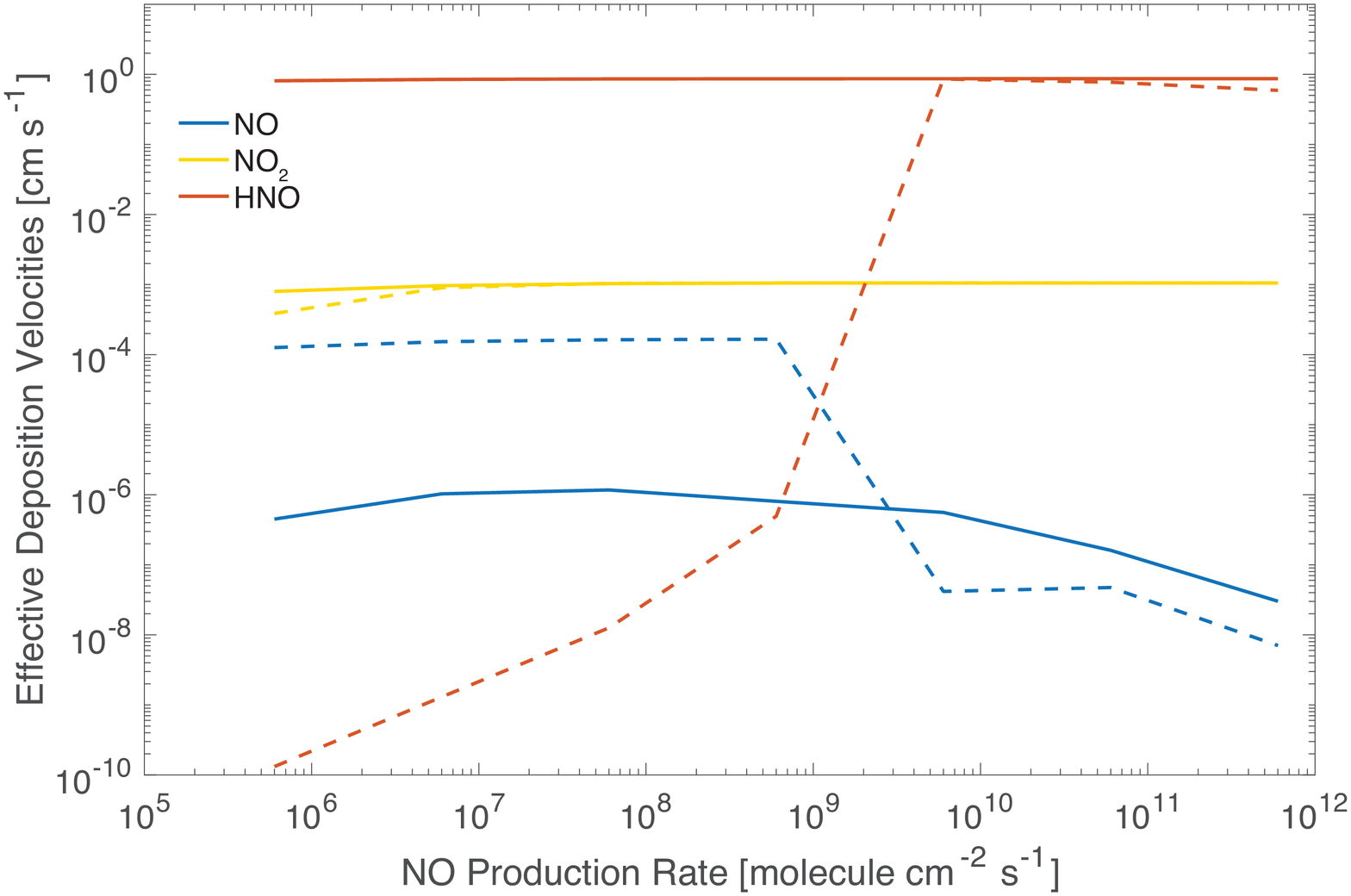}
    \caption{
    Effective deposition velocities in the converged atmosphere-ocean chemistry solutions. Solid lines are from the Sun-like star cases, and dashed lines are from the GJ~876 cases. The effective deposition velocities are self-consistently calculated, and are different from case to case.}
    \label{fig:vdep_real}
\end{figure}

The lifetime of molecular nitrogen in planetary atmospheres in contact with a liquid-water ocean for varied \ce{NO} fluxes from lightning and other energetic processes is shown in Figure \ref{fig:lifetime}. The lifetime is calculated from the deposition fluxes of \ce{NO}, \ce{NO2}, \ce{NO3}, \ce{N2O5}, \ce{HNO2}, \ce{HNO3}, and \ce{HNO4}, from the converged atmosphere-ocean chemistry solutions. The atmospheric abundances of these species are shown in Figures \ref{fig:sun_detail} and \ref{fig:gj876_detail}. The deposition flux of \ce{HNO} is not included in the calculation of the lifetime, as it is returned to the atmosphere in the form of \ce{N2O} (Section \ref{sec:fate}). With the effective deposition velocities calculated self-consistently from the ocean-chemistry models (Figure \ref{fig:vdep_real}), the deposition fluxes of weakly soluble species (\ce{NO} and \ce{NO2}) represent how fast the ocean can process them. 

The lifetime of molecular nitrogen is well longer than 1 billion years unless the NO flux is $>100$ times larger than the present-day Earth's lightning production rate. Interestingly, we see that the lifetime of nitrogen on planets around Sun-like stars is longer than that on planets around M dwarf stars. For instance, the lifetime under the lightning rate of present-day Earth is $\sim2$ billion years on an M dwarf's habitable planet, and that on a Sun-like star's habitable planet is 4-order-of-magnitude longer.

The atmospheric nitrogen chemistry is substantially modified with the inclusion of the oceanic feedback, i.e, the inability to deposit \ce{NO} and the return flux of \ce{N2O}. For a Sun-like star as the parent star, the atmosphere is always poor in \ce{O2} (Figure \ref{fig:sun_detail}), and thus oxidizing \ce{NO} is difficult. For a higher \ce{NO} production rate, the steady-state mixing ratios of \ce{NO}, \ce{NO2}, and \ce{HNO} increase, and so is the return flux of \ce{N2O}. The steady-state mixing ratio of \ce{N2O} thus also increases. The upper limit of the \ce{N2O} mixing ratio obtained from our model is $\sim10^{-8}$, still much smaller than that in present-day Earth's atmosphere ($\sim3\times10^{-7}$). The dominant form of nitrogen deposition is \ce{HNO3} when the \ce{NO} flux is $\le6\times10^8$ molecule cm$^{-2}$ s$^{-1}$, and it becomes \ce{HNO2} when the \ce{NO} flux is $\ge6\times10^9$ molecule cm$^{-2}$ s$^{-1}$. The surface abundance and thus the deposition rate of \ce{HNO} is larger than \ce{HNO2} and \ce{HNO3} -- it is however not counted as a net loss of atmospheric nitrogen. The steady-state mixing ratio of \ce{NO} can accumulate to a quite high level, and this is made possible by its very small effective deposition velocity (Figure \ref{fig:vdep_real}). In other words, the ocean cannot process the \ce{NO} so quickly. For the same reason, even a large surface abundance \ce{NO} does not imply a major deposition pathway.

The situation is more complex when the parent star is an M dwarf. Because M dwarfs emit strongly in the far-ultraviolet bandpass but weakly in the near-ultraviolet bandpass, their rocky planets in the habitable zone tend to accumulate \ce{O2} from photolysis of \ce{CO2} \citep{TIAN2014,domagal2014abiotic,Harman2015}. The \ce{NO}-\ce{NO2} catalytic cycle initiated by lightning cannot remove the photochemical \ce{O2} on an M dwarf's planet either (Hu et al. 2019, ApJ, submitted). Here we find the same phenomenon of abiotic \ce{O2} accumulation, and the exact amount of \ce{O2} has to do with the assumed \ce{NO} flux from lightning \citep[][and Hu et al. 2019, ApJ, submitted]{Harman2018}. The accumulation of abiotic \ce{O2} is not the focus of this paper, but the availability of free oxygen does impact the nitrogen chemistry and greatly reduces the lifetime of \ce{N2}. With the free oxygen, the atmosphere has up to 10 ppm of \ce{O3} in the stratosphere and is thus able to efficiently oxidize \ce{NO} via
\reaction{NO + O3 -> NO2 + O2. \label{NOO3}}
Compared to the Sun-like star cases, the M star cases have higher abundances of \ce{NO2}, \ce{NO3}, and \ce{HNO2} at the steady state. The higher abundance of \ce{NO2} also helps deposition of \ce{NO} via Reaction (\ref{R:NO}). \ce{HNO} is practically not produced in the atmosphere unless the NO flux from lightning is $\ge6\times10^9$ molecule cm$^{-2}$ s$^{-1}$. When it is produced, the corresponding return flux of \ce{N2O} can drive the atmospheric \ce{N2O} to up to $3\times10^{-8}$. To compare, the terrestrial (biological) emission rate of \ce{N2O} would lead to a much higher abundance of $\sim10^{-6}$ \citep[][]{Segura2005}. The response of \ce{HNO} and \ce{N2O} to an increasing lightning rate is not monotonic, and this reflects the competing effects of free oxygen, and a low level of near-ultraviolet irradiation and low abundances of \ce{OH} and \ce{HO2} in the atmosphere.

\section{Discussion} \label{sec:discussion}

\subsection{Lifetime of Nitrogen on Archean Earth}

We can apply the results to Archean Earth as the modeled atmosphere irradiated by a Sun-like star has an oxidation state similar to Earth before the rise of oxygen. Except for bolide impact that concentrated in the earliest time \citep{Mckay1988}, the production rate of \ce{NO} from lightning and hot volcanic vents would be in the range of $6\times10^7\sim6\times10^8$ molecule cm$^{-2}$ s$^{-1}$ \citep{Mather2004,Wong2017,Harman2018}. With this input, we find that the total flux of nitrogen deposition would in the range of $1.6\times10^4\sim1.5\times10^5$ molecule cm$^{-2}$ s$^{-1}$. In other words, only $\sim0.03\%$ of the reactive nitrogen produced in the atmosphere is permanently lost to the ocean. The lifetime of nitrogen is $10^4$ billion years or larger, implying that the \ce{N2} atmosphere is stable without any help from nitrate-consuming microbes.

Of the deposition flux of nitrogen species, approximately 80\% is \ce{HNO3} and 20\% is \ce{HNO2}. The flux of nitrate deposition we calculate is consistent in the ballpark with \cite{Wong2017} but we clarify the oceanic feedback to the gas deposition and we remove \ce{HNO} from effective deposition. Assuming that the residence time of this nitrite and nitrate is determined by the ocean cycling through high-temperature hydrothermal vents \citep[$\sim0.4$ billion years,][]{Wong2017}, and an average ocean depth of $\sim3$ km, we estimate the concentration of nitrate to be 0.9 -- 9 $\mu$M, and that of nitrite to be 0.2 -- 2 $\mu$M in the Archean ocean. If circulation through all hydrothermal vents causes the removal of nitrite and nitrate \citep{laneuville2018earth}, the residence time reduces to $\sim10$ million years and the nitrate and nitrite concentrations further reduce by two orders of magnitude.

Cycling through hydrothermal vents is probably not the only way to remove nitrite and nitrate in the ocean. \cite{ranjan2019nitrogen} compares the kinetic loss rate of oceanic nitrite and nitrate due to hydrothermal vents, ultraviolet photolysis \citep{zafiriou1974sources,carpenter2015chemistry}, and reactions with reduced iron \citep{jones2015stable,buchwald2016constraining,grabb2017dual,stanton2018nitrous}. The loss rates due to photolysis and reactions with reduced iron can be greater than that due to hydrothermal vents by orders of magnitude. This implies that the concentrations of nitrite and nitrate we estimate in this section is an upper limit and the actual concentrations can be much lower.

\subsection{Abiotic \ce{N2O} in Anoxic Atmospheres}

In this work we show that \ce{HNO} produced in the atmosphere would become \ce{N2O} when an aqueous environment exists. One might ask if this source of \ce{N2O} constitutes a ``false positive'' for using \ce{N2O} as a biosignature gas \citep[e.g.,][]{DesMarais2002}. With the coupled atmosphere-ocean model, we find that the abundance of \ce{N2O} produced by \ce{HNO} dehydrative dimerization is always smaller than the abundance of \ce{N2O} that would be produced from a source strength of current Earth's biosphere, by more than one order of magnitude, but it can be comparable to a lower biological \ce{N2O} production in Earth's anoxic past \citep[e.g.][]{rugheimer2018spectra}. This is true for either a Sun-like star or an M star as the parent star. The difference in the \ce{N2O} mixing ratio by more than one order of magnitude causes an appreciable difference in the \ce{N2O} spectral features in the infrared \citep[e.g.][]{rugheimer2018spectra}. The use of \ce{N2O} as a biosignature gas thus requires the detection of its source strength at the level of current Earth's biosphere.
%
%

\section{Conclusion}
\label{sec:conclusion}

We present a coupled atmosphere-ocean chemistry model to study the lifetime of molecular nitrogen (\ce{N2}) in planetary atmospheres in contact with a liquid-water ocean. The question of lifetime exists because nitrogen is the background gas for canonical planetary habitability scenarios and because nitrogen could be sequestered in the ocean when it is chemically converted to soluble compounds like nitrites and nitrates.

We clarify several important features of nitrogen's aqueous-phase chemistry for planetary applications. First, we find that dehydrative dimerization is the main loss pathway of \ce{HNO}, the dominant nitrogen species produced in anoxic atmospheres. This reaction produces \ce{N2O}, which is then released to the atmosphere and photodissociated to become \ce{N2}. This finding corrects the long-standing assumption that the \ce{HNO} would eventually become nitrate in the ocean. Second, we find that the deposition flux of \ce{NO} is always very small under anoxic conditions. These findings collectively indicate that sequestering nitrogen in the ocean requires atmospheric oxidation to at least as oxidized as \ce{NO2} and \ce{HNO2}.  

We determine that the lifetime of molecular nitrogen is well longer than 1 billion years unless the NO flux is $>100$ times larger than the present-day Earth's lightning production rate. As such, \ce{N2} atmospheres on Archean Earth and habitable exoplanets of both Sun-like and M dwarf stars are kinetically stable against aqueous-phase sequestration. This result affirms the nitrogen-based habitability on rocky planets.

\acknowledgments
This research was supported by NASA's Exoplanets Research Program grant \#80NM0018F0612. The research was carried out at the Jet Propulsion Laboratory, California Institute of Technology, under a contract with the National Aeronautics and Space Administration.

\end{document}